\documentclass[aps,prl,print,groupedaddress]{revtex4-2}

\begin{document}


\title{A Fundamental Equation for Gravitational Wave and Its Analogues in Type D Spacetimes}


\author{Zhong-Heng Li}
\email[]{zhli@zjut.edu.cn}
\affiliation{College of Science, Zhejiang University of Technology, Hangzhou 310032, China}

\date{\today}

\begin{abstract}
It is well known that Teukolsky equation of gravitational perturbations provides a powerful tool for the investigation of the ringdown phase of a binary black hole merger, but it is applicable only for final configurations in the Kerr cases. Here we introduce a new concept: the spin-coefficient connection. Employing it and a new transformation function, we find that the gravitational perturbations can be written as a single equation. Each coefficients of this equation has explicit physical significance. More importantly, the equation is universally applied to virtually any black hole since it does not depend on specific metrics and coordinate system. We also find that any massless field of nonzero spin $s<2$ obeys the gravitational wave-like equation, which shows that the fields can be used as the analogue models of gravitational wave.
\end{abstract}


\maketitle

\section{}
Up to now, all gravitational wave signal observed come from binary star systems [1-3], though the bulk is from binary black hole mergers. The evolution theory of binary systems will provide a powerful tool for studying astrophysical processes involving highly curved spacetime regions. The evolution of a black hole binary may be divided roughly into three stages: inspiral, merger, and ringdown [4,5]. During the early stages of an inspiral, one can resort to approximate solution to the problem using post-Newtonian they. When enter the merger phase the post-Newtonian theory will lose validity and should be taken over by numerical relativity. During ringdown phase after merger described using the perturbation theory of black holes.

The perturbation theory of black holes began to develop in 1957 under the influence of Regge and Wheeler's work [6]. So far one have done extensive studies in the field [7-14]. Not only is black-hole perturbation content rich, it is also technically difficult. In 1972, Teukolsky [15,16], using the Newman-Penrose formalism [17], succeeded in disentangling the perturbations of the Kerr metric, and wrote a master equation, which provide a powerful tool for investigating the evolution of the ringdown phase. Since in this stage, the system can be approximated by the evolution of perturbations of the final Kerr spacetime [18,19]. It is worth pointing out that the Teukolsky equation is also a powerful formula when one study extreme mass ratio inspirals.

As already mentioned, the Teukolsky equation is very useful for gravitational wave astronomy. However, the equation is only suitable for the Kerr metric, not other ones. In fact, there are various black holes in universe, or, to it another way, the results of binary star mergers are not just the Kerr black hole. On the other hand, directly from the Teukolsky equation itself, we cannot know how the background geometry affects gravitational waves. If there is a perturbation equation, which includes geometric quantities, and applies to various types of black holes, then it is very important for black hole perturbation theory. The purpose of this work is to investigate whether this is so.

In order to achieve the goal in a rigorous and general manner, we shall use the Newman-Penrose formalism, which is a tetrad formalism based on a set of four null vectors. The tetrad consists of two real null vectors, $l_{\mu}$ and $n_{\mu}$, and a pair of complex null vectors, $m_{\mu}$ and $\bar{m}_{\mu}$, which satisfies the orthonormal conditions, $l_{\mu}n^{\mu}=-m_{\mu}\bar{m}^{\mu}=1$, and $l_{\mu}l^{\mu}=n_{\mu}n^{\mu}=m_{\mu}m^{\mu}=\bar{m}_{\mu}\bar{m}^{\mu}=0$. The indexes are raised and lowered using global metric $g_{\mu\nu}$, which in terms of null vectors can be expressed as, $g_{\mu\nu}=2l_{(\mu}n_{\nu)}-2m_{(\mu}\bar{m}_{\nu)}$. Also, the metric can be written more compactly: $g^{\mu\nu}=\eta^{ij}\lambda^{\mu}\,_{i}\lambda^{\nu}\,_{j}$, where $i$ is the tetrad index, which are raised and lowered using
the flat metric $\eta_{ij}$ , and where $\lambda^{\mu}\,_{i}$ is defined by [17]
\begin{eqnarray}\label{1}
\lambda^{\mu}\,_{i}=(l^{\mu}, n^{\mu}, m^{\mu}, \bar{m}^{\mu}),
\end{eqnarray}
with
\begin{equation}\label{2}
\begin{array}{ll}
\eta^{ij}=\left(\begin{array}{lll}
0\,\,\,\,\,\,1\,\,\,\,\,\,0\,\,\,\,\,\,0\\
1\,\,\,\,\,\,0\,\,\,\,\,\,0\,\,\,\,\,\,0\\
0\,\,\,\,\,\,0\,\,\,\,\,\,0\,\,-1\\
0\,\,\,\,\,\,0\,\,-1\,\,\,\,\,\,0
\end{array}
\right).
\end{array}
\end{equation}

In the Newman-Penrose formalism, there are twelve spin coefficients which are represented by the symbols $\kappa, \varepsilon, \pi, \rho, \alpha, \lambda, \sigma, \beta, \mu, \tau, \gamma$ and $\nu$. Many of spin coefficients have direct geometric significance [20]. For instance, the vanishing of $\kappa$ is the condition for the integral curves of $l^{\mu}$ to be geodesic, while, if $\sigma$ is also zero, this congruence of geodesics is shear free. The same role is played by $\nu$ and $\lambda$ for the $n^{\mu}$-congruence. If the spin coefficients, $\kappa, \sigma, \lambda$ and $\nu$ vanish, we can conclude on the basis of the Goldberg-sachs theorem [21] that the Weyl scalars, $\psi_0, \psi_1, \psi_3 ,\text{and}\ \psi_4 $ must vanish in the chosen basis. The Weyl scalar $\psi_2$ does not, however, vanish. The spacetime with these properties have to be algebraically special, that is of Petrov type D [22,23]. In this study we focus on the universal equation of black hole perturbation theory. Therefore, backgrounds have to be the type D spacetime, since the black hole solutions of general relativity are all of Petrov type D [20].

In the Newman-Penrose formalism, the ten independent components of Weyl tensor are completely determined by the five complex Weyl scalars. All of the Weyl scalars have explicit physical meanings [24]. The gravitational quantities of interest will be $\psi_{0}$ and $\psi_{4}$, which are transverse components propagating in the $n^{\mu}$ and $l^{\mu}$ directions, respectively. For type D spacetimes, the $n^{\mu}$ and $l^{\mu}$ are both aligned with the two repeated principal null directions. The perturbation quantities $\psi_{0}^{B}$ and $\psi_{4}^{B}$ of $\psi_{0}$ and $\psi_{4}$ are invariant under gauge transformations and infinitesimal tetrad rotations [16], and are therefore completely measurable physical quantities, which represent the gravitational waves of spin weight 2 and spin weight -2, respectively.

Generally, the gravitational wave equation is not accurately separable, but in all of the type D metrics, it can be decoupled in the case of perturbations, and the equation for $\psi_{0}^{B}$ and $\psi_{4}^{B}$ take the forms [16]
\begin{eqnarray}\label{3}
[(D-3\varepsilon+\bar{\varepsilon}-4\rho-\bar{\rho})(\Delta-4\gamma+\mu)-(\delta+\bar{\pi}-\bar{\alpha}-3\beta-4\tau)(\bar{\delta}+\pi-4\alpha)-3\psi_2]\psi_{0}^{B}
=\kappa_{2}T_{0},
\end{eqnarray}
\begin{eqnarray}\label{4}
[(\Delta+3\gamma-\bar{\gamma}+4\mu+\bar{\mu})(D+4\varepsilon-\rho)
-(\bar{\delta}-\bar{\tau}+\bar{\beta}+3\alpha+4\pi)(\delta-\tau+4\beta)
-3\psi_2]\psi_{4}^{B}=\kappa_{2}T_{4}.
\end{eqnarray}
where $D, \Delta,$ and $\delta$ are the directional derivatives defined by
$D=l^{\mu}\partial_{\mu}, \quad \Delta=n^{\mu}\partial_{\mu}$, and $\quad \delta=m^{\mu}\partial_{\mu}$, $T_{0}$ and $T_{4}$ are the source terms [16], and $\kappa_{2}$ is the constant from the Einstein field equations. We follow the convention of using  $s$ to denote the spin, and $p$ denotes the spin weight (note that $p=\pm s$), then the perturbation quantities, the source terms and the constant can be rewritten as $\psi_{s-p}^{B}$, $T_{s-p}$, and $\kappa_{s}$, respectively.

To simplify the gravitational wave Eqs. (3) and (4), we introduce a transformation function $\Xi^{p-s}$, where $\Xi$ defined by
\begin{eqnarray}\label{5}
D\ln\Xi=-\rho, \quad \Delta\ln\Xi=\mu, \quad \delta\ln\Xi=-\tau, \quad \bar{\delta}\ln\Xi=\pi.
\end{eqnarray}
The assumption that $\psi_{s-p} ^{B}$ can be written in the form
\begin{eqnarray}\label{6}
\psi_{s-p} ^{B}=\Xi^{p-s}\Phi_{p}.
\end{eqnarray}
Putting this into Eqs. (3) and (4), we find that the two equations can be combined into a single statement:
\begin{eqnarray}\label{7}
&&\big\{[D-(2p-1)\varepsilon+\bar{\varepsilon}-2p\rho-\bar{\rho}](\Delta-2p\gamma+\mu)
-[\delta+\bar{\pi}-\bar{\alpha}-(2p-1)\beta-2p\tau](\bar{\delta}+\pi-2p\alpha)\nonumber\\
&&-(2p-1)(p-1)\psi_2\big\}\Phi_{p}=\kappa_{s}T_{p},
\end{eqnarray}
where $T_{p}=T_{s-p}/\Xi^{p-s}$.

To get a deeper insight into the Physical and geometrical properties hidden in Eq. (7), we introduce a new connection $\Gamma_{\mu}$, which we call the spin-coefficient connection, defined by
\begin{eqnarray}\label{8}
\Gamma_{\mu}=2\lambda_{\mu i}S^{i},
\end{eqnarray}
where
\begin{eqnarray}\label{9}
S^{i}=(-\gamma,\,\, -\varepsilon-\rho,\,\, \alpha,\,\, \beta+\tau).
\end{eqnarray}
Here $S^{i}$ is that we use the spin coefficients to construct a vector. Using the orthonormal conditions of the null vectors, namely $\lambda^{\mu}\,_{i}\lambda_{\mu j}=\eta_{ij}$, the inner product $\Gamma^{\mu}\Gamma_{\mu}$ is easy to calculate:
\begin{eqnarray}\label{10}
\Gamma^{\mu}\Gamma_{\mu}=8[\gamma(\varepsilon+\rho)-\alpha(\beta+\tau)].
\end{eqnarray}

Using the spin-coefficient connection (8), and Newman-Penrose equation [17], the gravitational wave Eq. (7) can be expressed in terms of the Weyl scalar $\psi_{2}$, and the Ricci scalar $R$:
\begin{equation}\label{11}
[(\nabla^{\mu}+p\Gamma^{\mu})(\nabla_{\mu}+p\Gamma_{\mu})-4p^{2}\psi_{2}+\frac{1}{6}R]\Phi_{p}=2\kappa_{s}T_{p}.
\end{equation}
Here $\nabla_{\mu}$ denotes the covariant derivative in the metric $g_{\mu\nu}$. Note that the spin-coefficient connection contains the spin coefficient $\rho$. According to the geometrical interpretation of $\rho$, the real and imaginary parts of $\rho$ are, respectively(minus), the expansion and the twist of the congruence of integral curves of $l^{\mu}$. Therefore, the first term within the square bracket in Eq. (11) tells us the information that there exist the contraction and rotatin when the wave functions evolve from one point to another. on the other hand, Penrose [25] point out that the effect of Weyl tensor is as a purely astigmatic lens, whereas the Ricci scalar is proportional to a cosmological constant for usual black hole solutions [26]. A positive cosmological constant provides a repulsive  contribution to gravitational effect. Hence, roughly speaking, the second and third terms within the bracket in Eq. (11) will lead to the convergency and divergence of wave functions in propagation processes.

The behaviour of gravitational waves in type D spacetimes is described by wave functions $\psi_{s-p}^{B}$ which are determined by Eq. (11). $\psi_{4}^{B}$ is particular important, since for outgoing waves at infinity, it is related to the metric perturbation by $\psi_{4}^{B}(r\rightarrow\infty)=-\partial_{t}^{2}(h_{+}-ih_{\times})$, and encodes the two polarization states of a gravitational wave [18].

Consider now massless fields for low spin $s=1/2, 1$, and $3/2$, which obey Weyl equation, Maxwell equation and Rarita-Schwinger equation, respectively. In type D spacetimes, these equation can be decoupled into [16,27]
\begin{eqnarray}\label{12}
[(D+\bar{\varepsilon}-\rho-\bar{\rho})(\Delta-\gamma+\mu)
-(\delta-\bar{\alpha}-\tau+\bar{\pi})(\bar{\delta}-\alpha+\pi)]\chi_{+1/2}^{(1/2)}=0,
\end{eqnarray}
\begin{eqnarray}\label{13}
[(\Delta-\bar{\gamma}+\mu+\bar{\mu})(D+\varepsilon-\rho)
-(\bar{\delta}+\bar{\beta}+\pi-\bar{\tau})(\delta+\beta-\tau)]\chi_{-1/2}^{(1/2)}=0.
\end{eqnarray}
\begin{eqnarray}\label{14}
[(D-\varepsilon+\bar{\varepsilon}-2\rho-\bar{\rho})(\Delta-2\gamma+\mu)
-(\delta+\bar{\pi}-\bar{\alpha}-\beta-2\tau)(\bar{\delta}+\pi-2\alpha)]\chi_{+1}^{(1)}=\kappa_{1}J_{+1}^{(1)},
\end{eqnarray}
\begin{eqnarray}\label{15}
[(\Delta+\gamma-\bar{\gamma}+2\mu+\bar{\mu})(D+2\varepsilon-\rho)
-(\bar{\delta}-\bar{\tau}+\bar{\beta}+\alpha+2\pi)(\delta-\tau+2\beta)]\chi_{-1}^{(1)}=\kappa_{1}J_{-1}^{(1)}.
\end{eqnarray}
\begin{eqnarray}\label{16}
[(D-2\varepsilon+\bar{\varepsilon}-3\rho-\bar{\rho})(\Delta-3\gamma+\mu)
-(\delta+\bar{\pi}-\bar{\alpha}-2\beta-3\tau)(\bar{\delta}+\pi-3\alpha)
-\psi_2]\chi_{+3/2}^{(3/2)}=0,
\end{eqnarray}
\begin{eqnarray}\label{17}
[(\Delta+2\gamma-\bar{\gamma}+3\mu+\bar{\mu})(D+3\varepsilon-\rho)
-(\bar{\delta}-\bar{\tau}+\bar{\beta}+2\alpha+3\pi)(\delta-\tau+3\beta)
-\psi_2]\chi_{-3/2}^{(3/2)}=0.
\end{eqnarray}
For the sake of simplicity, here we use $\chi_{p}^{(s)}$ and $J_{p}^{(s)}$ to represent wave functions and source terms, respectively. The wave equations listed above are are rather different, but if using the transformations $\chi_{p}^{(s)}=\Xi^{p-s}\Phi_{p}$, and $J_{p}^{(s)}=\Xi^{p-s}T_{p}$, [remember, $\Xi$ is determined by Eq. (5)], all the equations reduce to the form of Eq. (7). It follows inexorably that Eq. (11) describes all the perturbations, with the nonzero spins $s\leq2$.

For the source free case, Eq. (11) is taken to be
\begin{equation}\label{18}
[(\nabla^{\mu}+p\Gamma^{\mu})(\nabla_{\mu}+p\Gamma_{\mu})-4p^{2}\psi_{2}+\frac{1}{6}R]\Phi_{p}=0.
\end{equation}
It is surprising that the massless free-field equations for the nonzero spins $s\leq2$ have such a similar structure in virtually any black-hole spacetime. All the field equations collapse to a unique gravitational equation, which shows that the solutions have the common behavior, differing only in the value of the spin $s$. Hence these fields can be used to simulate gravitational waves, or, in other words, they are gravitational wave analogues.

For the massless free-field Eq. (18), there are several specific examples [28-30], which have only the form of first two terms in the square bracket. As an application of our theory, using Eqs. (5) and (8), one easily obtain the transformation functions and the spin-coefficient connections given by these examples.

We introduced a new concept, the spin-coefficient connection, $\Gamma_{\mu}$. Employing this and the new transformation function $\Xi^{p-s}$, we found that the gravitational perturbations could be written as a single equation in terms of the Weyl scalar $\psi_{2}$ and the Rcici scalar $R$, that is Eq. (11). This is the fundamental result of gravitational perturbation theory; as a practical matter, it may well be the very important equation in the final stages of merging black holes, since it is suitable for virtually any black hole.

Is is well-known that, in the weak field approximation, the Einstein field equation can be reduced to Maxwell like equations [31], which show the existence of gravitational effects similar to the electromagnetic ones. Here we found that Eq. (11) also governs the massless fields of arbitrary nonzero spin $s\leq2$. Note that the conclusion does not depend the strength of gravitational fields. For the source free case, Eq. (11) becomes (18), which shows that the fields can be used as the analogue models of gravitational wave in type D spacetime. This provides a sound basis for simulations of gravitational wave. For example, the electromagnetic waves in analogue spacetimes [32] can be used to mimic gravitational wave propagation in curved spacetimes.

Obviously, when $p=0$, the equation is just the (conformally invariant) massless scalar field equation. This is effectively the extension of Eq. (11) to zero value of spin $s$. Notice that the scalar field cannot be regarded as gravitational wave analogue, because it does not include the spin-coefficient connection $\Gamma_{\mu}$ and Weyl scalar $\psi_{2}$ terms.

Thus, all the massless fields of spin $s\leq2$ obey a single equation in type D spacetimes. Each coefficient in the equation has direct geometric significance. The coefficients can lead to the rotation, convergency and divergence of the wave functions in propagation processes. It is worth noting that our results not only can study gravitational waves and its analogues, but also can deal with other astrophysical applications, as quasinormal modes, scattering, and stability, etc.

\begin{acknowledgments}
This work was supported by the National Natural Science Foundation
of China under Grant No. 11475148.
\end{acknowledgments}

\subsection{}
\subsubsection{}

\bibliography{basename of .bib file}

\begin{thebibliography}{999}
\bibitem{1}B. P. Abbott \textsl{et al}. (LIGO Scientific and Virgo Collaborations), Phys. Rev. Lett. $\mathbf{116}$, 061102 (2016).
\bibitem{2}B. P. Abbott \textsl{et al}. (LIGO Scientific and Virgo Collaborations), Phys. Rev. Lett. $\mathbf{119}$, 161101 (2017).
\bibitem{3}B. P. Abbott \textsl{et al}. (LIGO Scientific and Virgo Collaborations), Phys. Rev. Lett. $\mathbf{125}$, 101102 (2020).
\bibitem{4}E. E. Flanagan and S. A. Hughes, Phys. Rev. D $\mathbf{57}$, 4535 (1998).
\bibitem{5}S. A. Hughes,  Annu. Rev. Astron. Astrophys. $\mathbf{47}$, 107 (2009).
\bibitem{6}T. Regge and J. A. Wheeler, Phys. Rev. $\mathbf{108}$, 1063 (1957).
\bibitem{7}R. A. Konoplya and A. Zhidenko, Rev. Mod. Phys. $\mathbf{83}$, 793 (2011).
\bibitem{8}Z. H. Li, Phys. Rev. D $\mathbf{62}$, 024001 (2000).
\bibitem{9}Z. H. Li, Classical Quantum Gravity $\mathbf{21}$, 1181 (2004).
\bibitem{10}Z. H. Li, Phys. Lett. B $\mathbf{643}$, 64 (2006).
\bibitem{11}Z. H. Li, Phys. Rev. D $\mathbf{80}$, 084013 (2009).
\bibitem{12}Z. H. Li, X. J. Wang, L. Q. Mi, and J. J. Du, Phys. Rev. D $\mathbf{95}$, 085017 (2017).
\bibitem{13}A. Zimmerman and Y. Chen, Phys. Rev. D $\mathbf{84}$, 084012 (2011).
\bibitem{14}L. C. B. Crispino, S. R. Dolan, and E. S. Oliveira, Phys. Rev. Lett. $\mathbf{102}$, 231103 (2009).
\bibitem{15}S. A. Teukolsky, Phys. Rev. Lett. $\mathbf{29}$, 1114 (1972).
\bibitem{16}S. A. Teukolsky, Astrophys. J. $\mathbf{185}$, 635 (1973).
\bibitem{17}E. Newman and R. Penrose, J. Math. Phys. (N.Y.) $\mathbf{3}$, 566 (1962)
\bibitem{18}J. Centrella, J. G. Baker, B. J. Kelly, and J. R. van Meter, Rev. Mod. Phys. $\mathbf{82}$, 3069 (2010).
\bibitem{19}I. Hinder, Classical Quantum Gravity $\mathbf{27}$, 114004 (2010).
\bibitem{20}S. Chandrasekhar, {\it The Mathematical Theory of Black Holes} (Oxford University Press, New York, 1983).
\bibitem{21}J. N. Goldberg and R. K. Sachs, Acta. Phys. Polonica, Supp. $\mathbf{22}$, 13 (1962).
\bibitem{22}W. Kinnersley, J. Math. Phys. (N.Y.) $\mathbf{10}$, 1195 (1969).
\bibitem{23}A. Z. Petrov, Gen. Relativ. Gravit. $\mathbf{32}$, 1665 (2000).
\bibitem{24}J. B. Griffiths and J Podolsk\'{y}, {\it Exact Space-Times in Einstein’s General Relativity} (Cambridge University Press, New York, 2009) .
\bibitem{25}R Penrose, {\it The Road To Reality: A Complete Guide to the Laws of the Universe} (Jonathan Cape, London, 2004).
\bibitem{26}J. B. Griffiths and J. Podolsk\'{y}, Int. J. Mod. Phys. D $\mathbf{15}$, 335 (2006).
\bibitem{27}G.F. Torres del Castillo, J. Math. Phys. (N.Y.) $\mathbf{30}$, 446 (1989).
\bibitem{28}D. Bini, C. Cherubini, R. T. Jantzen, and R. Ruffini, Prog. Theor. Phys. $\mathbf{107}$, 967 (2002).
\bibitem{29}D. Bini, C. Cherubini, R.T. Jantzen, and B. Mashhoon, Phys. Rev. D $\mathbf{67}$, 084013 (2003).
\bibitem{30}Z. Perj\'{e}s and M. Vas\'{u}th, Astrophy. J. $\mathbf{582}$, 342 (2003).
\bibitem{31}V. B. Braginsky, C. M. Caves, and K. S. Thorne, Phys. Rev. D $\mathbf{15}$, 2047 (1977).
\bibitem{32}C. Sheng, R. Bekenstein, H. Liu, S. Zhu, and M. Segev, Nat. Commun. $\mathbf{7}$, 10747 (2016).
\end{thebibliography}

\end{document}